\documentclass[preprint,12pt]{elsarticle}
\usepackage{amssymb,amsmath,amscd}
\usepackage{amsfonts,bm,latexsym}

\journal{Physics Letters B}

\def\be{\begin{equation}}
\def\ee{\end{equation}}
\def\bea{\begin{eqnarray}}
\def\eea{\end{eqnarray}}

\def\nn{\nonumber}
\def\p{\partial}
\def\Star{\,^{\star}\!}

\begin{document}
\begin{frontmatter}

\title{Thermodynamics of five-dimensional static three-charge STU black holes with squashed horizons}

\author{Shuang-Qing Wu\footnote{E-mail: sqwu@phy.ccnu.edu.cn, Corresponding author},
Dan Wen, Qing-Quan Jiang, and Shu-Zheng Yang}
\address{Institute of Theoretical Physics, College of Physics and Electric Information,
China West Normal University, Nanchong, Sichuan 637002, People's Republic of China}

\begin{abstract}
We present a new expression for the five-dimensional static Kaluza-Klein black hole solution with squashed $S^3$
horizons and three different charge parameters. This black hole solution belongs to $D = 5$ $N = 2$ supergravity
theory, its spacetime is locally asymptotically flat and has a spatial infinity $R \times S^1 \hookrightarrow S^2$.
The form of the solution is extraordinary simple and permits us very conveniently to calculate its conserved charges
by using the counterterm method. It is further shown that our thermodynamical quantities perfectly obey both the
differential and the integral first laws of black hole thermodynamics if the length of the compact extra-dimension
can be viewed as a thermodynamical variable.
\end{abstract}


\end{frontmatter}

\section{Introduction}

Five-dimensional black holes \cite{TI11} with squashed horizons are of special interest in Kaluza-Klein theory where
the fifth dimension is assumed to be compactified into a circle. A simplest example for this in the five-dimensional
Einstein-Maxwell theory is the static charged, squashed Kaluza-Klein black hole solution found by Ishihara and Matsuno
\cite{IM06} via applying the so-called squashing transformation to the five-dimensional Reissner-Nordstr\"{o}m black
hole solution. This black hole has horizon-topology of a $S^3$ sphere that is deformed by the squashing function. At
the infinity ($r\to r_\infty$), the spacetime approaches to the direct product of a flat time dimension and the
asymptotic structure of the self-dual Taub-NUT instanton with NUT charge $r_\infty/4$. In other words, the black
hole spacetime is asymptotically a twisted $S^1$ fibre bundle over $M^{1,3}$, which can be interpreted as a static
charged Reissner-Nordstr\"{o}m black hole sitting on the self-dual Taub-NUT instanton \cite{CT10}.

Sooner after the work done in Ref. \cite{IM06}, the squashing procedure was then successfully applied to generate new
black hole solutions with squashed horizons, due to its simplicity of the method. Subsequently, a large class of new
Kaluza-Klein black hole solutions in five dimensions had been constructed so far in vacuum Einstein gravity \cite{TW06},
Einstein-Maxwell-dilaton gravity \cite{SY06,NY13}, $D = 5$ minimal Einstein-Maxwell-Chern-Simons supergravity \cite{NIMT08}
and $U(1)^3$ supergravity \cite{ST10}. Later, several black hole solutions with squashed horizons in a five-dimensional
G\"{o}del universe had also been presented in Refs. \cite{MINT08,TI08,TIMN09}.

In recent years, there are also a lot of attention being paid to various aspects of squashed Kaluza-Klein black holes.
For example, thermodynamical properties \cite{CCO06,KI07,KI08,SSW08,SN09,PW10}, Hawking radiation \cite{NY13,IS07,CWS08,
WLLR10,MU11,HL10,HL11}, perturbation stability \cite{KMIS08,IKKMSZ08,NK10}, quasinormal modes \cite{HWCCL08,HWC09},
gravitational lens \cite{LCJ10,CLJ11,SBV13}, geodesic motion \cite{MI09} and Kerr/CFT correspondence \cite{PW10} have
been investigated for this kind of black holes in recent years. In particular, Cai {\em et al.}, \cite{CCO06} first
adopted the boundary counter-term method \cite{MS06,AR06} to investigate thermodynamics of the static charged, squashed
Kaluza-Klein black holes found in Ref. \cite{IM06} and showed that it is the counter-term mass, which is equal to the
Abott-Deser mass, rather than the Komar mass, that obeys the differential first law when the radius of the compact
extra-dimension is considered as a constant. The counter-term method was then frequently applied to calculate the boundary
stress-energy tensor and the conserved charges of a large class of Kaluza-Klein black holes \cite{TW06,SY06,CCO06,KI07,
SSW08,NY11} with squashed horizons. Moreover, thermodynamics of the five-dimensional squashed Reissner-Nordstr\"{o}m
black hole was investigated in details in Ref. \cite{KI08} where the counter-term mass is in agreement with the mass
calculated by the background subtraction method if the Kaluza-Klein monopole background is considered as a natural
reference spacetime for the squashed Kaluza-Klein black hole.

Our aim of this Letter is mainly concerned with thermodynamics of the three-charge squashed Kaluza-Klein black hole
solution to $D = 5$ $N = 2$ ungauged supergravity theory. A rotating solution in this theory was previously presented
in Ref. \cite{ST10} where the author obtained the solution by directly applying the squashing transformation to the
three-charge Cveti\v{c}-Youm black hole \cite{CY96} with two equal rotation parameters. Some thermodynamical quantities
were given in \cite{ST10}, however they can not consistently fulfil both the differential and the integral first laws
of squashed black hole thermodynamics. The reason for this is simply because the ADM mass computed there does not obey
the standard first law of thermodynamics for the squashed Kaluza-Klein black holes \cite{KI07,KI08}. What is more, the
contribution of three dipole charges should but had not been taken into consideration in Ref. \cite{ST10}, when the
rotation is included. Therefore, it is clear that thermodynamical properties of this solution had not been correctly
studied ever before in the previous literature, unlike the minimal case \cite{PW10}, and it deserves a deeper
investigation of its thermodynamics, which consists of the main subject of our work.

However, the rotating version of the three-charge squashed solution presented in Ref. \cite{ST10} is very complicated
after performing the coordinate transformations from ($t, r$) to ($\tau, \rho$) even if one only considers the static
case, so the expression for the solution is not suitable for our purpose to study its thermodynamical properties. For
the sake of simplicity, in this Letter we shall focus on the nonrotating case only. Our strategy is to seek another new
form for the three-charge static squashed black hole solution which is different from the one previously presented in
Ref. \cite{ST10}. To derive the solution, we have applied a triple-repeated lift-boost-reduction procedure \cite{EE03,
LMP10} to the five-dimensional static squashed Schwarzschild black hole solution which has a normalized Killing time
vector at spatial infinity. The derivation is essentially parallel to that of the static three-charge STU black hole
solution in Ref. \cite{HMS96}. The final expressions for the solution are much simpler than those presented in Ref.
\cite{ST10} because three gauge potentials and two dilaton scalar fields associated with it all vanish at the infinity.
For this form of the solution, one can very easily calculate its conserved mass and gravitational tension by using the
counterterm method, and show that all thermodynamical quantities computed for our three-charge static Kaluza-Klein
black hole with squashed $S^3$ horizons perfectly satisfy both the differential and the integral first laws of squashed
black hole thermodynamics if the length of the compact extra-dimension can viewed as a thermodynamical variable. When
three charges are set to be equal, our results completely reproduce those obtained in Ref. \cite{CCO06} after some
suitable identifications of solution parameters.

The remaining part of our Letter is organized as follows. In Sec. \ref{Sect2}, we first present a new form of the static
three-charge squashed black hole solution. Then, the boundary counterterm method is adopted to calculate the conserved
mass and gravitational tension which together with the entropy, horizon temperature, three charges and their corresponding
electrostatic potentials completely satisfy both the differential and the integral first laws when we consider the length
of the extra-dimension as a thermodynamical variable. Our Letter ends up with a summary of our work and the related future
plan.

\section{Static three-charge squashed black
hole solution and its thermodynamics} \label{Sect2}

In this section, we present a new simple form of the five-dimensional three-charge static squashed black hole solution
and investigate its thermodynamics. To generate the solution, we start from the static squashed Schwarzschild solution
after performing the appropriate coordinate transformations and use a thrice-repeated sequence \cite{EE03,LMP10} of
lifting to six dimensions, performing a Lorentz boost, and reducing again to $D = 5$. The final expressions for the
metric and three Abelian gauge potentials are concisely given by
\bea
ds^2 &=& (h_1h_2h_3)^{1/3}\Bigl[-\frac{1 -\rho_1/\rho}{h_1h_2h_3} d\tau^2
 +\frac{1 +\rho_0/\rho}{1 -\rho_1/\rho} d\rho^2 \nn \\
&& +\rho(\rho +\rho_0)(d\theta^2 +\sin^2\theta d\psi^2)
 +\frac{\rho_0(\rho_0 +\rho_1)}{1 +\rho_0/\rho}(d\phi +\cos\theta d\psi)^2 \Bigr] \, , \label{3cSqKK} \\
A_I &=& \frac{c_Is_I\rho_1}{h_I\rho}\,d\tau \, , \label{3cgp}
\eea
and three scalars are
\be
X_I = \frac{(h_1h_2h_3)^{1/3}}{h_I} \, , \qquad h_I = 1 +s_I^2\frac{\rho_1}{\rho} \, ,
\ee
where $c_I = \cosh\delta_I$, and $s_I = \sinh\delta_I$, in which $\delta_I$'s are three charge parameters. In the
solution, the coordinate $\rho$ varies from 0 to $\infty$, and ($\theta, \phi, \psi$) are three Eulerian angles,
taking the ranges $0 < \theta < \pi$, $0 < \psi < 2\pi$, $0 < \phi < 4\pi$.

At spatial infinity ($\rho \to \infty$), we have $h_I \to 1$ and $X_I \to 1$, so three gauge potentials $A_I$ tend
to zero and two dilaton scalar fields ($\varphi_1, \varphi_2$) behave asymptotically like
\bea
\varphi_1 = \frac{(s_1^2 +s_2^2 -2s_3^2)\rho_1}{\sqrt{6}\rho} +O(\rho^{-2}) \, , \qquad
\varphi_2 = \frac{(s_1^2 -s_2^2)\rho_1}{\sqrt{2}\rho} +O(\rho^{-2}) \, ,
\eea
while the metric (\ref{3cSqKK}) approaches to
\be
ds^2 = -d\tau^2 +d\rho^2 +\rho^2(d\theta^2 +\sin^2\theta d\psi^2) +\rho_0(\rho_0 +\rho_1)(d\phi +\cos\theta d\psi)^2 \, .
\ee

Just like the uncharged solution ($\delta_I = 0$), the asymptotic structure of our solution (\ref{3cSqKK}) is also a
four-dimensional flat Minkowski spacetime with a compact extra-dimension. That is, its spacetime is locally asymptotically
flat and has an asymptotic boundary topology $R \times S^1 \hookrightarrow S^2$, and $\p_\phi$ generates the twisted $S^1$
fibre bundle at spatial infinity, with a constant size $2\pi\sqrt{\rho_0(\rho_0+\rho_1)}$. The horizon topology is, however,
obviously a squashed $S^3$ sphere.

In the case where all three charges are set to equal ($s_I = s$), we find that by setting 
$$\tilde{\rho} = \rho +s^2\rho_1 \, , \quad \tilde{\rho}_0 = \rho_0 -s^2\rho_1 \, , \qquad
 \rho_+ = c^2\rho_1 \, , \quad \rho_- = s^2\rho_1 \, , $$
and rescaling the gauge potential by a factor $\sqrt{3}$, the above solution exactly reproduces the static squashed
Reissner-Nordstr\"{o}m black hole solution presented in Refs. \cite{IM06,CCO06}.

We have verified that our solution solves the full set of equations of motion derived from the Lagrangian of the $D = 5$,
$N = 2$ supergravity theory whose action is
\bea
\mathcal{I} &=& \frac{1}{16\pi G} \int_\mathcal{M} d^5x \bigg\{\sqrt{-g}\Big[R -\frac{1}{2}(\p\varphi_1)^2
 -\frac{1}{2}(\p\varphi_2)^2 -\sum\limits_{I=1}^3\frac{1}{4}X_I^{-2}F^I_{\mu\nu}F^{I\mu\nu}\Big] \nn \\
&& +\frac{1}{4}\varepsilon^{\mu\nu\alpha\beta\lambda}F^1_{\mu\nu}F^2_{\alpha\beta}A^3_{\lambda}\bigg\}
 +\frac{1}{8\pi G}\int_{\p\mathcal{M}} K\sqrt{-h} d^4x \, , \label{act}
\eea
where the Chern-Simons term is included for the completeness, but it makes no contribution to the action in the nonrotating
case. In the boundary Gibbons-Hawking term, $K$ is the trace of extrinsic curvature $K_{ij} = (n_{i ; j} +n_{j ; i})/2$ for
the boundary $\p\mathcal{M}$ with the induced metric $h_{ij}$, $R$ is the bulk scalar curvature, and $F_I = dA_I$ are strengths
associated to three $U(1)$'s gauge fields. Two dilaton scalar fields ($\varphi_1, \varphi_2$) are related to three scalars
$X_I$ by
\be
X_1 = e^{-\varphi_1/\sqrt{6} -\varphi_2/\sqrt{2}} \, , \quad
X_2 = e^{-\varphi_1/\sqrt{6} +\varphi_2/\sqrt{2}} \, , \quad
X_3 = e^{2\varphi_1/\sqrt{6}}\, .
\ee

It is now the position to investigate thermodynamical properties of our solution given above. An suitable approach for
this aim is to make use of the counterterm method \cite{MS06,AR06} to compute the conserved charges since our solution
is also an asymptotically flat spacetime with boundary topology $R \times S^1\hookrightarrow S^2$. In the method, a
counterterm, which is a functional only of the curvature invariants of the induced metric on the boundary, is added to
the boundary term at infinity, so that a regular gravitational action is obtained without any modification of the
equations of motion.

We consider the following simple counterterm proposed by Mann and Stelea \cite{MS06}
\be
I_{ct} = \frac{1}{8\pi G} \int d^4x\sqrt{-h}\sqrt{2\mathcal{R}} \, ,
\ee
where $\mathcal{R}$ is the Ricci scalar with respect to the boundary metric $h_{ij}$. Varying the action (\ref{act}) with
this counterterm leads to the boundary stress-energy tensor
\be
T_{ij} = \frac{1}{8\pi G}\Big[K_{ij} -Kh_{ij} -\Psi\big(\mathcal{R}_{ij} -\mathcal{R}h_{ij}\big)
 -h_{ij}h^{kl}\Psi_{; kl} +\Psi_{; ij} \Big] \, ,
\ee
where $\Psi = \sqrt{2/\mathcal{R}}$, and the covariant derivative is defined with respect to the induced metric $h_{ij}$ on
the boundary.

If the boundary geometry has an isometry generated by the Killing vector $\xi$, then $T_{ij}\xi^j$ is divergence free,
so the conserved charge associated with it is given by
\be
\mathcal{Q} = \int_\Sigma d^3 S^i \, T_{ij}\xi^j \, ,
\ee
which represents the conserved mass $M_{ct}$ in the case when $\xi = \p_\tau$, and the gravitational tension $\mathcal{T}$
if $\xi = \p_\phi$.

After some calculation, we find the needed components of the stress tensor as follows
\bea
T_{\tau\tau} &=& \frac{\rho_0 +\rho_1(2 +s_1^2 +s_2^2 +s_3^2)}{2\rho^2} +O(\rho^{-3}) \, , \\
T_{\phi\phi} &=& \frac{2\rho_0 +\rho_1}{2\rho^2} +O(\rho^{-3}) \, , \\
T_{\psi\psi} &=& \frac{\rho_0^2 +\rho_0\rho_1 -\rho_1^2}{8\rho^3} +O(\rho^{-4}) \, .
\eea
Then it is straightforward to calculate the conserved mass and gravitational tension as
\bea
M_{ct} &=& \pi\sqrt{\rho_0(\rho_0 +\rho_1)} \Big[\rho_0 +(2 +s_1^2 +s_2^2 +s_3^2)\rho_1 \Big] \, , \\
\mathcal{T} &=& \frac{\rho_0}{2} +\frac{\rho_1}{4} \, .
\eea

On the horizon, the entropy $S = A/4$ and temperature $T = \kappa/(2\pi)$ can be easily obtained as
\bea
S &=& 4\pi^2c_1c_2c_3\sqrt{\rho_0}\rho_1^{3/2}(\rho_0 +\rho_1) \, , \\
T &=& \frac{1}{4\pi c_1c_2c_3\sqrt{\rho_1(\rho_0 +\rho_1)}} \, ,
\eea
and three electrostatic potentials are given by
\be
\Phi_I = (A^I_\mu\chi^\mu) \big|_{\rho=\rho_1} = \frac{s_I}{c_I} \, .
\ee
Finally, we have three electric charges corresponding to the gauge potentials by completing the integral
\be
Q_I = \frac{1}{16\pi}\int_{S^3}X_I^{-2} \Star F_I
 = \pi c_Is_I\rho_1\sqrt{\rho_0(\rho_0 +\rho_1)} \, .
\ee

It is not difficult to verify that the above thermodynamical quantities completely satisfy both the different and the
integral first laws of black hole thermodynamics
\bea
dM_{ct} &=& TdS +\Phi_1dQ_1 +\Phi_2dQ_2 +\Phi_3dQ_3 +4\pi\mathcal{T}d\mathcal{L} \, , \\
M_{ct} &=& \frac{3}{2}TS +\Phi_1Q_1 +\Phi _2Q_2 +\Phi_3Q_3 +2\pi\mathcal{T}\mathcal{L} \, ,
\eea
where $2\pi\mathcal{L}$ is the length of the extra-dimension, and $\mathcal{L} = \sqrt{\rho_0(\rho_0 +\rho_1)}$ can be
identified with twice of the NUT charge. In the equal-charge case ($Q_i = Q/\sqrt{3}$), our results exactly recover
those obtained in Ref. \cite{CCO06} where the radius $2\pi\mathcal{L}$ of the extra-dimension was considered as a
constant. Here, we view it as a thermodynamical variable for the self-consistence of the integral first law.

\section{Conclusions}

In this Letter, we have presented a new form for the static three-charge STU black hole solution with squashed horizons.
The expressions for the metric, three gauge potentials and two scalar fields are rather simple and very convenient for
us to investigate its thermodynamics. By means of the counterterm method, all thermodynamical quantities of our solution
are easily calculated and have been shown to fulfil both differential and integral first laws of the squashed black hole
thermodynamics when the length of the extra-dimension is considered as a thermodynamical variable.

It is interesting to extend the present work to the rotating charged case \cite{ST10} where the form of the squashed black
hole solution is definitely inconvenient for studying its thermodynamics. A promising routine to resolve this difficulty
is to regenerate this solution from the neutral seed of the rotating Kaluza-Klein squashed black hole solution \cite{TW06}
with a Killing time vector normalized at spatial infinity.

\section*{Acknowledgements}
S.Q. Wu is supported by the National Natural Science Foundation of China (NSFC) under Grant Nos. 11275157 and 10975058.
Q.Q. Jiang and S.Z. Yang are, respectively, supported by the NSFC under Grant Nos. 11005086 and 11178018.

\end{document}